\documentclass[article]{aa}
\usepackage{natbib}
\usepackage{multirow}
\bibpunct{(}{)}{;}{a}{}{,}
 \usepackage[dvips]{graphicx}

\DeclareGraphicsExtensions{.ps}                      
\usepackage{txfonts}
\begin{document}
   \title{Simulation of planet detection with the SPHERE IFS}


   \author{D. Mesa\inst{1}, R. Gratton\inst{1}, A.Berton\inst{1}, J.Antichi\inst{2}, C.Verinaud\inst{2},
          A.Boccaletti\inst{3}, M. Kasper\inst{4}, R.U. Claudi\inst{1}, S. Desidera\inst{1}, E. Giro\inst{1}, 
          J.-L. Beuzit\inst{2}, K. Dohlen\inst{5}, M. Feldt\inst{6}, D. Mouillet\inst{2}, G. Chauvin\inst{2}, and
          A. Vigan\inst{5,7}
          }

   \institute{\inst{1}INAF-Osservatorio Astronomico di Padova, Vicolo dell'Osservatorio 5, Padova, ITALY, 35122-I \\
                   \inst{2}UJF-Grenoble 1 / CNRS-INSU, Institut de Planétologie et d'Astrophysique de Grenoble (IPAG) UMR 5274, Grenoble, F-38041, France\\
                   \inst{3}LESIA-Observatoire de Meudon, 5 place Jules Janssen,  92195 Meudon, France\\
                    \inst{4}European Southern Observatory, Karl-Schwarzschild-Strasse 2, D-85748 Garching, Germany\\
                     \inst{5}LAM, UMR 6110, CNRS, Universit\'e de Provence, 38 rue Fr\'ed\'eric Joliot-Curie, 13388 Marseille Cedex 13, France\\
                      \inst{6}Max Planck Institute for Astronomie, Konigsthul 17, Heidelberg, Germany\\
                   \inst{7}School of Physics, University of Exeter, Stocker Road, Exeter EX4 4QL, United-Kingdom}

   \date{Received  / accepted }


  \abstract
   {}
   {We present \textbf{simulations} of the perfomances of the future SPHERE IFS instrument designed for imaging 
extrasolar planets in the near infrared (Y, J, and H bands). }
   {We used the IDL package code for adaptive optics simulation (CAOS) to prepare a series of input point spread functions
(PSF). These feed an IDL tool (CSP) that we designed to simulate the datacube resulting from the SPHERE IFS.
We performed simulations under different conditions to evaluate the contrast that IFS will be able to reach and to verify the impact of physical propagation within the limits of the near field of the aperture approximation (i.e. Fresnel propagation). We 
then performed a series of simulations containing planet images to test the capability of our instrument to correctly classify 
the found objects. To this purpose we developed a separated IDL tool. }
   {We found that using the SPHERE IFS instrument and appropriate analysis techniques, such as multiple spectral differential
imaging (MDI), spectral deconvolution (SD), and angular differential imaging (ADI),  we should be able to image companion 
objects down to a luminosity contrast of $\sim 10^{-7}$ with respect to the central star in favorable cases. Spectral 
deconvolution resulted in the most effective method for reducing the speckle noise. We were then able to find most of the 
simulated planets (more than $90\%$ with the Y-J-mode and more than the $95\%$ with the Y-H-mode) for contrasts down to
$3\times 10^{-7}$ and separations between 0.3 and 1.0 arcsec.
The spectral classification is accurate but seems to be more precise for late T-type spectra than
for earlier spectral types. A possible degeneracy between early L-type companion objects and field objects (flat spectra)
is highlighted. The spectral classification seems to work better using the Y-H-mode than with the 
Y-J-mode.}
   {}


\titlerunning{IFS simulations}
\authorrunning{Mesa et al.}
   \maketitle
%

\section{Introduction}
A large  number of extrasolar planets have been discovered in  the last fifteen years through indirect methods such as 
radial velocities and transits . Although in the past few years some objects with planetary mass have been imaged around 
stellar and substellar objects like HR 8799~\citep{Ma08}, Fomalhaut~\citep{Kal08}, 2M 1207~\citep{Ch09}, and $\beta$
Pictoris~\citep{Lag10}, imaging of extrasolar planets is still very challenging because of the high planet vs star luminosity 
contrast ($10^{-6}$ for young giant planets and down to $10^{-8}$-$10^{-10}$ for old giant and rocky planets) and the small separation with respect to the central star (few tenths of arcsec for a planet at $\sim$10 AU at some tens of pc). \\
The next generation of instruments aimed at imaging extrasolar planets will exploit extreme adaptive optics (XAO) 
systems to correct aberrations up to a high order, providing a high Strehl ratio (SR) and high-efficiency coronagraphs to attenuate the on-axis PSF and reduce its diffraction pattern. The combination of these two devices should be able to reduce the stellar background down to a value of around $10^{-5}$ at separations of a few tenths of arcsec. The residual 
background will be given mainly by the speckle noise generated by the atmosphere and the telescope pupil-phase distortion. 
To further improve the contrast achievable with these instruments, it will be mandatory to apply differential imaging 
techniques, such as angular differential imaging (ADI) ~\citep{Ma06}, simultaneous spectral differential imaging (S-SDI) (see 
e.g. \citealt{Ma05}), and spectral deconvolution (SD) (see~\citealt{Th07}).\\
In the next years in particular, three instruments will be able to exploit these techniques to image extrasolar planets. These 
are the Gemini Planet Imager (GPI) at the Gemini South Telescope~\citep{Mac06}, SPHERE at the ESO Very Large
Telescope (VLT)~\citep{Beu06}, and Project 1640, which is already working at the 5 m Palomar telescope (see
\citealt{Crepp10}).\\
In particular, SPHERE will include three scientific channels: (i) a differential imager and dual band polarimeter called IRDIS 
that will operate in the near infrared between the band Y and $K_s$~\citep{Do08}; (ii) a polarimeter called ZIMPOL that will 
perform differential imaging exploiting the polarized light reflected from the planetary atmosphere in the visual 
band~\citep{Th08}; (iii) an integral field spectrograph (IFS) that will supply simultaneous images at different wavelengths in 
the near infrared between the Y and the H bands~\citep{Cl08}.\\
Integral field spectrographs also have the potential of providing the spectra of the detected faint companions at close
separation, thus allowing much better characterization. IFSs similar to the one designed for SPHERE are also present in GPI 
and in Project 1640, and are foreseen for future planet imagers like EPICS designed to work for the future E-ELT~\citep{Kasper10}.
In this paper we present the results obtained from the simulations we developed to evaluate and to optimize the 
performances of the SPHERE IFS.  In Section~\ref{IFS} we give a very short summary of the SPHERE IFS instrument, in 
Section~\ref{sim} we
describe the simulation tools that we used in our work, in Section~\ref{dataan} we describe the methods used for the data 
analysis of the output of our simulations, in Section~\ref{simres} we present the results of the various simulation runs, in 
Section~\ref{dataansof} we describe the software that we wrote for the IFS data analysis and the results obtained testing 
it on the output of our simulations, while in Section~\ref{conclusion} we report our conclusions.  


\section{SPHERE IFS description}\label{IFS}
The SPHERE IFS  is designed to work in two different wavelength ranges: (i) 0.95-1.35 $\mu$m (Y-J-mode) with a resolution of R=50 and (ii) 0.95-1.65 $\mu$m (Y-H-mode) with a resolution of R=30. These two ranges and resolutions are achieved through two different dispersers (two Amici prisms - see~\citealt{Oli00}). The IFS is composed of several
subsystems:
\begin{itemize}
\item the integral field unit (IFU)
\item the collimator optics system
\item a filter wheel
\item the disperser optics system
\item a camera optics system that can be moved to focus spectra on the detector or to produce dithering to reduce noise 
related to the flat fielding
\item a 2048$\times$2048 Hawaii II detector with pixel of 18 $\mu$m housed in a cryostat
\end{itemize}
The novel lenslet IFU concept upon which this spectrograph is based (BIGRE, \citealt{JA09}) allows the entrance slits plane to 
be made of images of the telescope focal plane and not of images of the telescope pupil, as ordered in the 
classical TIGER design~\citep{Ba95}. In this design, each lenslet is an afocal system with two powered surfaces. The 
thickness of the array is then given by the sum of the focal lengths of the lenslets of the two arrays. The main advantage of 
the BIGRE configuration over the TIGER one is that it allows a strong reduction of the cross-talk between adjacent lenslets as 
demostrated by~\citet{JA09}. The microlens array is composed of $145\times 145$ hexagonal 
lenslets with a pitch of 161.5 $\mu m$ (corresponding to $\sim$0.012 arcsec). Each lenslet is masked with a circular aperture
with a factor of 0.98 to avoid straylight. The full field of view (FOV) of the instrument is a square with a side of 1.77 arcsec. 
The total length of the whole instrument from the first surface of the IFU to the detector plane is 1061.89 mm. A 
more detailed description of the whole instrument can be found, e.g., in \citet{Cl10}.

\section{Simulation description}\label{sim}
\begin{figure}
\begin{center}
\includegraphics[width=8.0cm]{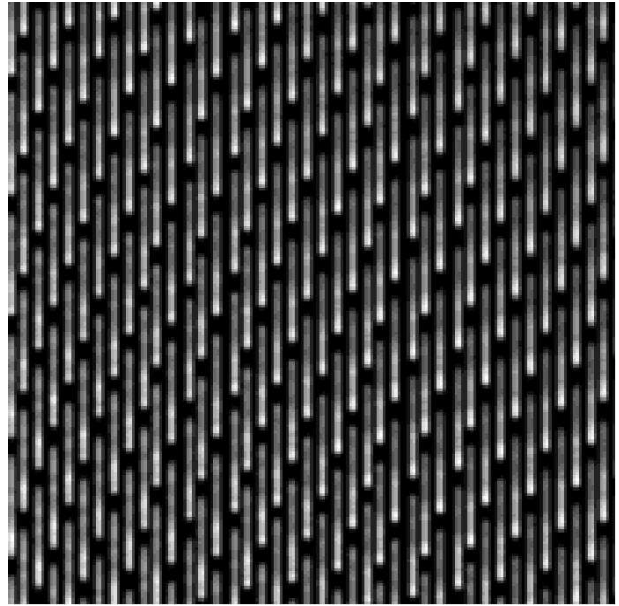}
\caption{A small portion of spectra resulting from one of the CSP simulations. \label{spettrini}}
\end{center}
\end{figure}

We exploited two software tools for our simulations:
\begin{itemize}
\item the SPHERE package of the CAOS software
\item the CSP code\footnote{CSP stands for CHEOPS Simulation Program. The code was originally developed in the
context of the first feasibility study for a direct imaging planet finder for VLT called CHEOPS~\citep{Feldt03}.}
\end{itemize}
The CAOS system~\citep{Car04} is an IDL based software that aims to simulate the behavior of a generic adaptive optic (AO) 
system from the atmospheric propagation of light to the sensing of the wavefront aberrations and the corrections through a deformable mirror. This is done with a Fraunhofer approach, so that it cannot be used to properly evaluate the impact of Fresnel propagation (see Section~\ref{fresnelimpact}). An end-to-end numerical tool has been developed for the 
simulation of the whole SPHERE instrument within the CAOS environment. It contains detailed instrumental modeling of the 
Extreme adaptive optics systems, of IRDIS and ZIMPOL~\citep{Car08}. A module simulating the SPHERE IFS has been also 
developed to properly take both the real and the imaginary parts of the image forming on the lenslet plane into account. In 
principle, this could allow a complete treatment of the cross-talk among the lenslets when studying the impact of light 
propagation through the BIGRE. However, the execution of this module turned out to be very time consuming so that it was 
not possible to use it for a large number of detailed simulations. To overcome this difficulty, we used a shorter code that 
calculates the impact of the cross-talk between adjacent lenslets (coherent) and  adjacent spectra (incoherent) by
providing the beam propagation over a sub-sample of 7 hexagonal lenslets. This code is described in detail in~\citet{JA09}. 
After running this code we concluded that a value of the cross-talk equal to or less than $10^{-2}$ was completely adequate 
for meeting the objectives of our instrument.\\
We then decided to use our (IDL oriented) code called CSP to perform all  the simulations of light propagation within the 
IFS, while we decided to use the SPHERE CAOS package to provide real intensities over the IFU entrance focal plane as input 
for CSP. For this, we performed simulations using the CAOS IRDIS module with 100 atmospheric phase screens at 64 
different wavelengths ranging between 0.95 and 1.35 $\mu m$ (or between 0.95 and 1.65 $\mu m$ in the Y-H-mode case).  
There are enough atmospheric screens is large enough to ensure that static speckles dominate noise, as expected in real 
cases, and to ensure that the PSF has an overall shape representing a realistic stellar halo.\\
\begin{figure}
\begin{center}
\includegraphics[width=8.0cm]{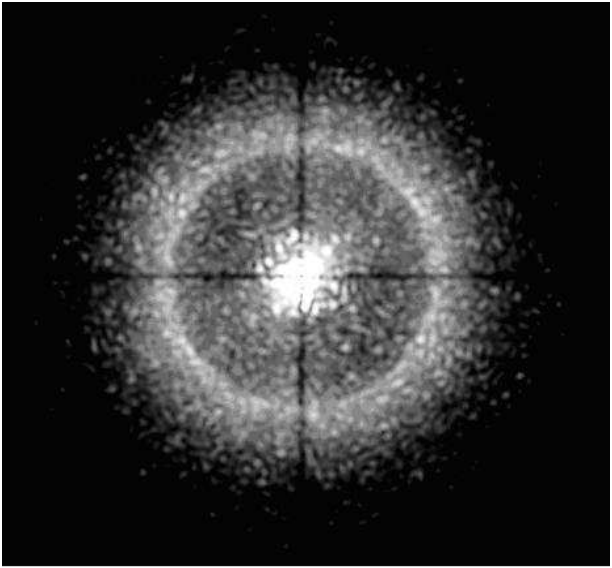}
\caption{Monochromatic PSF resulting from the CAOS simulation. The bright corona corresponds to the
outer working angle of the XAO system. Its radius is roughly 0.5 arcsec at the working wavelength. This image has been
obtained using a 4-quadrant coronagraph. The cross structure centered on the center of the image is the signature of this
type of coronagraph. \label{inputpsf}}
\end{center}
\end{figure}
In Figure~\ref{inputpsf} we display a monochromatic PSF obtained from the CAOS simulations. Even if the SPHERE 
package of the CAOS system allows simulating different types of coronagraphs, we preferred to use only a 
4-quadrant one for our simulations~\citep{Boccaletti08}. In this way, however, our results are still representative as we were 
not interested in investigating the performances of all the SPHERE instrument coronagraphs. The choice of making a 
simulation with only 64 wavelengths was determined because for more wavelengths, the program 
saturates our computer memory. However, CSP requires 269 PSFs at different wavelengths as input and to obtain them, we
performed interpolations starting from the ones resulting from the CAOS simulation\footnote{Note, however, that 64 
wavelengths are enough to properly sample both the expected spectral extension of speckles over the whole FOV
(hypersampling - see~\citealt{JA09}) and each pixel with approximately two sampling per pixel. This should guarantee
that interpolation errors are under control.}.\\
An early version of the CSP code, described in~\citet{Be06}, has been deeply modified to take 
variations in the instrument optical design into account.  CSP only considers the real part of the image on the lenslet 
plane and then propagates it through the IFS spectrograph using a Fraunhofer approach, but it can include a treatment of the 
cross-talk through a parametric approach. The code can be divided into different parts:
\begin{itemize}
\item The image formation part simulates the propagation of the light through the instrument and its main goal is to produce a
final image with all the spectra. For each spectral step,  the exact number of photons passing through every
microlens is calculated, as well as the correct position projected on the detector of the center of each microimage. The 
intermediate images generated by this process are then convolved with a microlens PSF prepared in advance. All the 
monochromatic images are properly shifted to account for the spectral dispersion due to the Amici prisms and are
then summed up to create the spectra. Finally the code adds noises to the image: Poisson noise and all the detector noises.
An example of the output of this part of the code is given in Figure~\ref{spettrini}.
\item The calibration part performs the same procedure as is described at the previous point using a monochromatic uniform 
illumination of the IFU (and not a PSF) as input. This part simulates the wavelength calibration lamps and is performed at 
three different wavelengths. The code, then, reads from the images resulting from these procedures the spectra (having a 
template that indicates the position of the spectra at the minimum wavelength). Every spectrum is fitted with a Gaussian curve 
(using the IDL routine GAUSSFIT) and the code finds the center of the Gaussian and its error. Finally, through an 
appropriate interpolation, the code calculates the shift for each lenslets, using the positions of the previuosly calculated three centers and the theoretical position of the center (well known because we know the wavelength of the calibration lamp). At 
the end, these results are saved in a wavelength map file where at every pixel of the image is associated a well-defined 
wavelength. 
\item The last part of the simulation procedure is to create of the datacube with the monochromatic images that will
be the instrument final output. To this aim we derive a rectangular grid from the original hexagonal pattern of the IFU. 
This is done by creating a square grid of points at the same distance from each other.  For every wavelength, the flux value 
associated to every point of the grid is calculated by considering the three nearest points at the given wavelength as 
calculated in the calibration step of the procedure and saved in the wavelength map. The calculation is made with a mean of 
the fluxes of these three points weighted according to the distance from the grid point considered.    
\end{itemize}

Where not specified, the simulations were performed assuming a G0 spectral type central star with an absolute magnitude 
of J=3.75 and at a distance of 10 pc from the Sun. A total exposure time of 1 hour was generally 
simulated even if, in some cases, longer exposure times were simulated. For these simulations we assumed a readout noise 
of 10 $e{-}$, a dark current of 0.1  $e{-}$, and a flat field error of $10^{-4}$ (hereinafter detector noise).

\section{Data analysis methods}\label{dataan}
High-performance coronagraphs within an extreme AO system like the one adopted in SPHERE, which gets its sampling 
frequency equal to 20 cycles/pupil, allows imaging of companion objects down to a contrast of $10^{-5}$ within the whole 
FOV of its IFS (2.5 arcsec diagonal), except for separations smaller than $\sim$0.1 arcsec from the central star.  However, to 
fulfill the goal of the SPHERE instrument to image giant planets around young nearby star, contrasts of $\sim 10^{-6}-10^{-7}$ are requested. To this aim, speckle noise has to be reduced by a further factor of 10 - 100. This
is done by applying some differential imaging analysis techniques to the final datacube extracted from the IFS data, such as 
the simultaneous spectral differential imaging (S-SDI)~\citep{Ma00} and the spectral deconvolution 
(SD)~\citep{Th07}. A natural evolution of the S-SDI technique, using an IFS, has been defined during our work and is 
described in detail in Section~\ref{mdi}. Another possible technique, normally applied with other analysis 
techniques, is angular differential imaging (ADI)~\citep{Ma06}.\\
In this section we briefly present the algorithms we used to implement these methods. 

\subsection{Multiple differential imaging}\label{mdi}
As previously said, this techique is an extension to more spectral channels of the previous S-SDI techniques, such as the single differential imaging for two channels and the double differential imaging for three channels~\citep{Ma00}. 
The final result of the CSP simulation code consists of a datacube composed of 33 (for the Y-J-mode) or 38 (for the Y-H-mode) 
monochromatic images. On these images we apply the following steps:
\begin{enumerate}
\item The images are divided into two groups: planetary images (monochromatic images at wavelengths where the planet 
signal is potentially present) and reference images (monochromatic images at wavelengths where the planet signal is very 
weak or absent) according to giant-planet atmosphere models.
\item We then distinguish two different cases:
\begin{itemize}
\item Single differences:
\begin{itemize}
\item A reference image is assigned to each planetary image.
\item For each pair, the reference image is spatially scaled (through an interpolation procedure) to the planetary image
according to the wavelength ratio between the wavelengths of planetary and reference image.
\item The scaled reference image is subtracted from the planetary one.
\end{itemize}
\item Double differences:
\begin{itemize}
\item Two reference images are assigned to each planetary image, with the wavelength respectively shorter and longer 
than the planetary one. The two reference images are chosen in such a way that their wavelength separations
from the image containing the planet signal are the same.
\item For each group of three images, the reference images are spatially scaled to the planetary image according to the 
wavelength ratio between the wavelengths of planetary and reference images.
\item The three images are combined according to the double-difference formula defined by~\citet{Ma00}.
\end{itemize}
\end{itemize}
\item The procedure at step 2 should eliminate most of the speckle pattern. If the pairs are selected so that the planet 
image is only present in one of the two images, the planet will not be canceled out.
\item A weighted average of the cleaned differential images provides the best final result suitable for the planet
search. We adopted a weight for each single differential image, which is the reciprocal of the wavelength difference 
between the two (or three) images subtracted to obtain the considered one. In this way we give a greater weight to differences between images with a smaller wavelength separation, where the speckle pattern has a stronger correlation. 
Since the planetary images are not scaled, the planet position will not shift with wavelength.
\end{enumerate}
There are three critical issues in this procedure:
\begin{enumerate}
\item To properly work we have to make an assumption about  the spectra of the companion objects that we are looking 
for. Moreover, this works much better for spectra with large absorption bands such as for methane-dominated planets.
\item Each interpolation introduces noise. In our approach, the number of interpolations is
effectively reduced to only one per pair (two for each group of three images when using the double differential imaging).
\item Pairing of monochromatic images, and optimal weighting should be given according to the main noise source:
\begin{itemize}
\item If errors are dominated by photon noise, the best procedure is to assign the same weights to all pairs. In this 
case, pairs should be selected to have similar (or even constant) wavelength separations.
\item If errors are dominated by calibration errors (speckle residuals), the best procedure in single differential imaging is to
create pairs having the smallest possible wavelength separation, compatible with the gradients present in the planetary
spectra. In this case weights should be assigned according to the inverse of the square of wavelength separation.
\item For what concerns double differences, this last approach is limited by the intrinsic width of the emission peaks in the
planetary spectrum. Practically, we expect a very small advantage by creating groups of three images with the smallest 
possible wavelength differences. It should then be more advantageous to have various  groups of three images with the 
same wavelength difference and give the same weight to all of them.
\end{itemize} 
\end{enumerate}

\subsection{Spectral deconvolution} \label{specdeconv}
\begin{figure}
\begin{center}
\includegraphics[width=8.0cm]{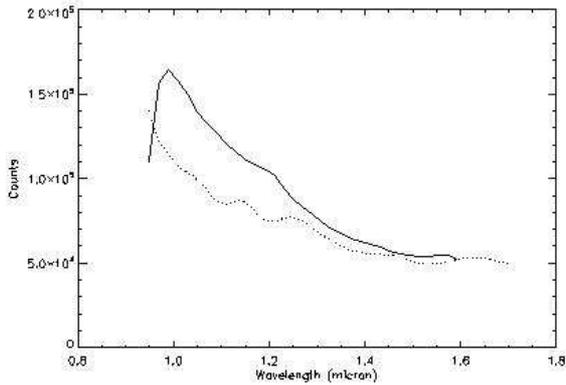}
\caption{Example of the simulated spectra on a spaxel rescaled according to the wavelength for SPHERE IFS simulations.
The dotted line represents the value of one single pixel on the PSFs used as input for the CSP code. The solid line
represents the same thing on the images of the datacube resulting from the CSP simulation (passage of the light 
through the instrument and correct calibration to create the datacube). The absolute values of these two curve have been 
normalized to be able to overplot one on the other.
\label{specdeconv2}}
\end{center}
\end{figure}

This method was proposed for the first time in~\citet{Sparks02} and further developed in~\citet{Th07}. It exploits  
that speckles are expected to change regularly with wavelength. Outside a given separation, defined as the bifurcation 
radius (BR),  the speckle spatial excursion over the spectral range is larger than the planet size so that the speckle pattern 
associated to the star can be reconstructed and eliminated using regions unaffected by the planet image. Differently from 
the MDI described in the previous section, no assumption about the spectra of the companion objects is needed.\\
Spectral deconvolution should offer some advantage over the differential imaging approach, at least outside the
BR, because it uses the companion spectrum as a whole. The value of the the SPHERE IFS BR is
around 0.20 arcsec for the Y-J-mode and about 0.12  arcsec for the Y-H-mode. The procedure we followed is 
composed of four steps:
\begin{itemize}
\item We scaled single images provided by the CSP data extraction algorithm to a reference wavelength (in this case we 
chose the central wavelength among those of the 33 or 38 monochromatic images). Because of this rescaling, the planet
will be in different positions in every image.
\item We plotted the spectrum for every spaxel of the rescaled datacube (see Figure~\ref{specdeconv2}) and calculated
a polynomial fitting function using $1/\lambda$ as independent variable. The polynomial degree depends on the distance from 
the center of the image, in units of the BR. The value of this fitting function is then subtracted from every 
spectrum. The fit allows the modulation of a given stellar halo speckle brightness with wavelength to be taken into 
account but its degree is small enough not to fit a potential planetary signal. This should eliminate, or at least reduce, the 
speckles or diffraction residuals.
\item The subtracted images are then rescaled back to the original scale according to their wavelength, in order to
mantain fixed the planet position in all of them.
\item To search for planet signal, the three-dimensional datacube is collapsed to a bi-dimensional image given by the
cross-correlation of the spectra in each spaxel with a template planet spectrum. This procedure enhances the signal-to noise 
of the final image. In general, in our simulations we use a methane-dominated spectrum. However, as we see in 
Section~\ref{dataansof}, this procedure works well with either a flat or an L-type spectrum, too.
\end{itemize}

\subsection{Angular differential imaging}\label{adi}
In general, we assume that observations are done with the field fixed with respect to the IFU. In this case, the pupil rotates 
with time on an alt-az telescope, a typical value being $30^{\circ}$ over a 1 hour exposure time. In this framework\footnote{
Angular differential imaging would actually be better applied by keeping the pupil static and leaving the field to rotate with
respect to the detector.}, angular differential imaging (ADI) can be applied to reduce the speckle noise further. Various codes 
have been written to perform ADI on real images (see~\citealt{Ma06a}). Here, we considered a variant of this method 
that we defined as azimuthal filtering (''azimuthal'', meaning along arcs at a constant radius). This procedure is composed 
of the following steps:
\begin{itemize}
\item For each given pixel, we searched for all spaxels that have similar separation (distance from the center). In our
procedure, the annulus width was set at 1 pixel.
\item We plotted the value of the intensity at the selected wavelength for each of these spaxels against the azimuth angle.
\item We drew a fitting line through these points using a cubic spline curve through the averages of these points
within arcs of length $4\lambda /D$. After various tests, we chose this value  to avoid canceling the 
planet signal.
\item We subtracted the intensity value on the fitting line from the intensity at the selected wavelength in that spaxel.
\item The procedure was then repeated for all wavelengths.
\item The procedure was iterated over all spaxels.
\end{itemize}
While this procedure does not completely eliminate the impact of static speckles, it also works well  for quasi-static
speckles, which are speckles having a lifetime longer than field rotation but shorter than the total exposure time.

\section{Simulations results}\label{simres}
\begin{figure}
\begin{center}
\includegraphics[width=8.0cm]{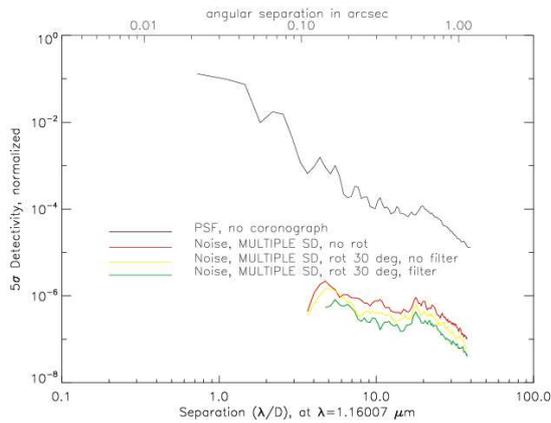}
\caption{Run of the $5\sigma$ calibration limit with separation for three cases: (i) no field rotation (red line), rotation by 
$30^{\circ}$ with no azimuthal filtering apllied (yellow line), rotation by $30^{\circ}$ with azimuthal filtering applied (green
line). \label{rotation1}}
\end{center}
\end{figure}

\begin{figure}
\begin{center}
\includegraphics[width=8.0cm]{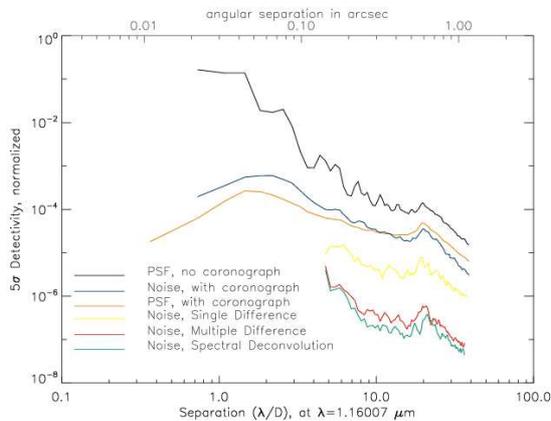}
\caption{Run of the $5\sigma$ calibration limit with separation for a very bright star. The case shown is for $30^{\circ}$
field rotation with azimuthal filtering. Detector noises and a cross-talk with a value of $10^{-2}$ were introduced too. 
Red line is the result obtained with multiple differential imaging while the green is with the spectrum deconvolution method.}
\label{specdeconv4}
\end{center}
\end{figure}

In this section we review the most important results obtained from our simulations. As said in previous sections, we expect a significant improvement in the contrast using the MDI method compared to a simple S-SDI, when 
exploiting all of the many monochromatic images provided by an IFS. In particular we expect 
that the contrast scales with the square root of the number of independent single differences that we can realize when using 
the whole spectrum. A further improvement can be obtained by correctly coupling  the images at different wavelengths. 
Since the contrast scales with the wavelength separation, we can pair monochromatic images in ascending order of 
wavelength separation and weight them in descending order according to their wavelength separation.\\
\begin{figure}
\begin{center}
\includegraphics[width=8.0cm]{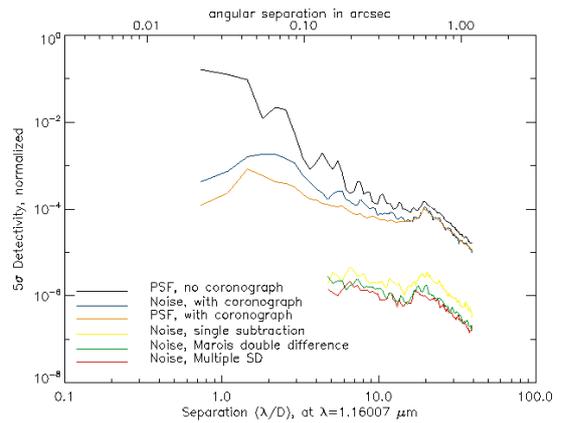}
\caption{Comparison between $5\sigma$ contrast obtained with single difference (yellow line), with multiple single
differences (red line), and with multiple double differences (green line) from IFS simulations. These results were obtained
for a simulation where no detector noise, no cross-talk, and no rotation were considered. \label{compplot}}
\end{center}
\end{figure}
In Figure~\ref{compplot} we display these results for a simulation where no detector noise, no cross-talk, and no 
rotation were considered. In particular,  we can see that no further gain is instead obtained using the multiple double-difference method. This is mainly because realistic double differences should be made using a rather large 
wavelength separation, because of the intrinsic width of the methane bands. In this simulation, as in all the following
ones, the jump in the plots around $20\lambda /D$ is given by the coronagraph outer working angle effects. \\
We can further improve the contrast obtained with our instrument by exploiting the rotation of the field with respect to the 
pupil. As seen in Figure~\ref{rotation1}, the improvement is better at large separations, as expected because more noise 
realizations can be sampled. If quasi-static speckles dominate, the results improve with the 
square root of the angle (and of the separation), since the planet images sample different noise realizations while rotating 
around the stellar image. In this case the azimuthal filtering procedure (described above in Section~\ref{adi}) can be applied.
\\
SD should provide better results than MDI, at least for separations larger than
the BR. This is confirmed by the plots displayed in Figure~\ref{specdeconv4} where we show the run of the
$5\sigma$ calibration limit for a very bright star. The case shown is for $30^{\circ}$ field rotation with azimuthal filtering.
In this case we introduced the detector noises using the values indicated at the end of Section~\ref{sim} and a cross-talk 
total amount of $10^{-2}$, too. Results obtained with the spectral deconvolution are slightly better than those obtained 
using the multiple differential imaging, with difference on the order of 0.2 dex ($\sim$ 0.5 mag). As expected, better results 
are obtained when the Y-H-mode is considered. In this second case the difference is on the order of 0.3 dex ($\sim$ 0.7-0.8 
mag), and the gain is appreciable even at a small separations (0.15 arcsec).\\
\begin{figure}
\begin{center}
\includegraphics[width=8.0cm]{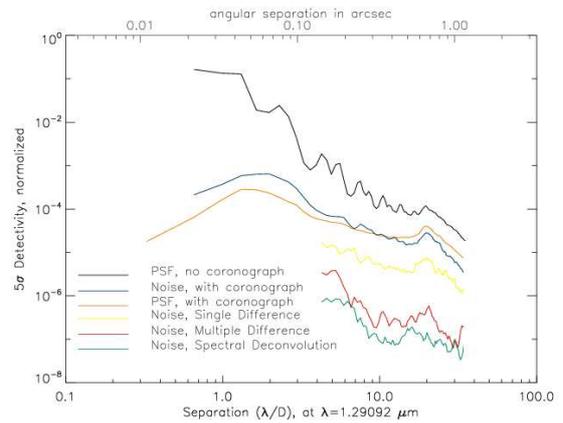}
\caption{Same as Figure~\ref{specdeconv4} but for the Y-H-mode. \label{specdeconv5}}
\end{center}
\end{figure}
From all these plots we can see that, by using SPHERE IFS and an appropriate combination of the analysis methods described 
above, we should be able to reach contrasts on the order of $10^{-7}$ or even better at large separations from the central 
star.

\begin{table}
\caption{$5\sigma$ calibration limit (30 degree field rotation, azimuthal filtering). In this table MDI stays for multiple
differential imaging, while SD stays for spectral deconvolution. \label{tabsummary}}
\begin{center}
\begin{tabular}{c c c c c}
\hline\hline
          &                            &   \multicolumn{3}{c}{Sep. (arcsec)}   \\  \cline{3-5}
Mode  &   Analysis   &  \multicolumn{1}{c}{0.15 }  &  0.5    &   1.0    \\  \hline
\multirow{2}{*}{Y-J-}  &   \multirow{2}{*}{MDI}  & $2.06\times 10^{-6}$   &  $3.09\times 10^{-7}$   &  $9.89\times 10^{-8}$   \\   
    &       &   (14.21 mag) & (16.27 mag)  & (17.51 mag)  \\ 
  &  \multirow{2}{*}{SD}  &  Inside BR  &  $1.64\times 10^{-7}$  &  $7.68\times 10^{-8}$   \\   
  &   &   &  (16.96 mag)  &  (17.79 mag) \\ \hline
\multirow{2}{*}{Y-H-}  &  \multirow{2}{*}{MDI}  &  $1.61\times 10^{-6}$  &  $1.87\times 10^{-7}$   &  $1.87\times 10^{-7}$    \\   
  &    &  (14.48 mag)  &   (16.82 mag)  &   (16.82 mag)  \\   
   &   \multirow{2}{*}{SD} &  $5.85\times 10^{-7}$   &  $1.13\times 10^{-8}$   &  $1.55\times 10^{-7}$   \\  
&  &   (15.58 mag)  &   (17.37 mag)   &    (17 .03 mag)  \\ \hline
\end{tabular}
\end{center}
\end{table}
A synthesis of the results  from our simulations is presented in Table~\ref{tabsummary} where we listed the contrasts 
obtained at different separations from the central star using the two analysis methods for Y-J and Y-H-modes. 

\subsection{Impact of Fresnel propagation}\label{fresnelimpact}
Out-of-pupil optics could have a strong impact on the performances of any differential technique adopted in high-contrast 
imaging due to Fresnel propagation, as described by~\citet{Ma06}. An optic that is not conjugated to a pupil 
plane will modify the light distribution in a chromatic way because at this location the beam intensity distribution depends on 
wavelength through diffraction effects. The closer the optic is to a focal plane, the larger this chromaticity. Even more
severe is the fact that this chromaticity is no longer smooth, but cyclic along the spectrum, when the optic is conjugated to
a height that is several times the Talbot length defined as
\begin{equation}\label{talbot}
L_T=2\Lambda^2/\lambda
\end{equation} 
where $\lambda$ is the light wavelength and $\Lambda$ the period of a single sinusoidal component of the wavefront across the pupil. For an aberration with a given period, the pupil complex amplitude presenting the electromagnetic field changes from a pure wavefront error to a pure amplitude error over a quarter of the Talbot length. Since the Talbot length is 
different for different periods, a decorrelation occurs that depends on angular separation. The farther an optic is  from the pupil plane (in multiples of Talbot length), the more the decorrelation along spectral domain will be, and speckle correlation 
will be broken.\\
In the case of SPHERE, the Talbot effect was expected to be strong for those optical components located before the lenslet
array, such as the entrance window, the ADC, the derotator and the coronagraphic mask \citep{Yai10}.\\
To evaluate the impact of the Fresnel propagation, we cannot use the CAOS package that is based on the Fraunhofer
propagation. We then exploited the PROPER code~\citep{Kr07} to create new PSFs that are then used as 
input for the CSP code. In Table~\ref{tabfresnelparameter} we list the parameters used to calculate the Fresnel propagation
in all the simulations. We report the values of the conjugated distance and the wavefront error (WFE) rms for all the 
considered optical surfaces. 
\begin{table}
\begin{center}
\caption{Most important parameters used for the Fresnel propagation calculation. \label{tabfresnelparameter}}
\begin{tabular}{c c c}
\hline\hline
Opt. surface  &  Con. Distance (km)   &   rms WFE (nm)  \\  \hline
DTTS	&   414   &   5   \\
Collimator  &   396   &   15 \\
Mirror 2  &   1492   &  10   \\
Mirror 3  &   4440   &   10  \\
Field lens  &   10722   &   15  \\  \hline
\end{tabular}
\end{center}
\end{table}
To save computing time, we performed all these simulations without considering the effects of the atmosphere (using only 1 atmospheric phase screen). Of course, this is not realistic, because it yields contrasts that are to optimistics. However, the 
comparison is still meaningful for evaluating the impact of Fresnel propagation itself.\\

\begin{figure}
\begin{center}
\includegraphics[width=8.0cm]{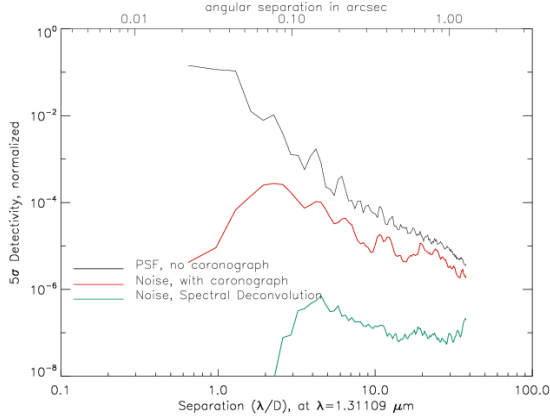}
\caption{Plot resulting from a simulation without the Fresnel propagation and without any rotation of the FOV (Y-H-mode).
\label{nofresnel}}
\end{center}
\end{figure}
\begin{figure}
\begin{center}
\includegraphics[width=8.0cm]{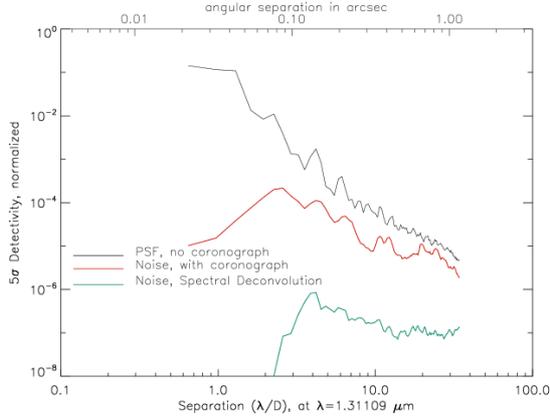}
\caption{Same as Figure~\ref{nofresnel} but with the Fresnel propagation. \label{fresnel}}
\end{center}
\end{figure}

From comparison of the plots in Figs.~\ref{nofresnel} and \ref{fresnel} resulting from simulations that do not include
and that do include the Fresnel propagation effects, respectively, one can see that the differences between the two cases 
are very small, as confirmed by the data reported in Table~\ref{tabdatafresnel} where the values of the contrast at different 
separations for simulations performed not considering (second column) and considering (third column) the Fresnel 
propagation are compared. From the results of these simulations, we can then conclude that the effects of the Fresnel 
propagation are not too great for our instrument. This is because of the large conjugated distances of the optics 
(see Table~\ref{tabfresnelparameter}) and the not very large pupil size. Fresnel propagation is much more a concern for 
extremely large telescopes, like E-ELT (see~\citealt{Antichi10}).
\begin{table}
\begin{center}
\caption{Values of the contrast at different separations for simulations with and without accounting for Fresnel 
propagation. \label{tabdatafresnel}}
\begin{tabular}{c c c}
\hline \hline
Sep. (arcsec)   &   No Fresnel     &    Fresnel   \\   \hline
0.2     &    $2.40\times 10^{-7}$    &   $3.50\times 10^{-7}$       \\
0.4     &   $8.35\times 10^{-8}$     &   $1.31\times 10^{-7}$       \\
0.6     &   $1.17\times 10^{-7}$     &   $1.23\times 10^{-7}$       \\
0.8     &   $9.80\times 10^{-8}$     &   $1.12\times 10^{-7}$      \\
1.0     &   $9.76\times 10^{-8}$     &   $1.19\times 10^{-7}$       \\  \hline
\end{tabular}
\end{center}
\end{table}

\section{Data analysis software for companion detection and spectral classification}\label{dataansof}
Through the simulations described in the previous sections, we have demonstrated the capability of the SPHERE IFS 
instrument to image extrasolar planets down to a contrast of $\sim 10^{-7}$ at a separation of a few tenths of arcsec. 
However, to fully characterize the newly discovered planets  (temperature, chemical composition of the atmosphere, 
etc.), it is important to be able to reconstruct their spectra at a high-fidelity level. \\
To test this capability, we prepared a pipeline for a data analysis of the calibrated datacube resulting from our
simulations. This procedure is composed of five different steps:
\begin{enumerate}
\item Speckle noise subtraction from the original datacube using the spectral deconvolution algorithm.
\item Sum of all the resulting images to create a single multi-wavelength image.
\item Search for companion objects on the summed image.
\item Extraction of a spectrum for every object found.
\item Spectral classification of every object.
\end{enumerate}
The simulation was performed using the same PSFs as in the previous simulations. The FOV is rotated by 
$30^{\circ}$ during the observation and the central star is a G0V star at a distance of 10 pc (this corresponds to a magnitude J=3.75). Every
simulation contains five planets in different positions but at the same separation from the central star (the planets in the same
simulation are identical). We perfomed simulations with planets at a separation of 0.3, 0.5 and 1.0 arcsec from the central 
star. To avoid overlapping of the planets PSFs at the smallest separation,  we decided to replace these
single simulations with three different ones containing two different planets each (for this reason only for the case at 0.3 
arcsec we have six planets for every single case
and not five). We then performed different simulations with different luminosity contrasts between the planets and the central 
star, and adopted values of $10^{-5}$, $3\times 10^{-6}$, $10^{-6}$, and $3\times 10^{-7}$. Finally, we performed
different simulations with different input spectra: we used a late type T-dwarf spectrum (T7), an early type T-dwarf spectrum
(T2), a late type L-dwarf spectrum (L8), and an early type L-dwarf spectrum (L0), taken from the spectra libraries described
in Section~\ref{procdesc}. To test the capability of our procedure to distinguish between a companion object and a 
background star, we performed simulations using  a flat spectrum (M2) at the low resolution of our instrument as input.
Indeed, at this resolution, all the stellar spectra are expected to be flat. Moreover, we did not include 
any faint background galaxies because they are expected to be spatially resolved as extended objects by our instrument.\\
While all these spectra come from objects in the solar neighborhood (old objects), our results do not lose generality
because, as shown in Section~\ref{gravity}, the detectability of the companion objects at a fixed effective temperature 
is not determined by the gravity effects, which in turn is the main difference between young and older substellar objects.

\subsection{Procedure description}\label{procdesc}
In this section we will describe our reduction procedure in more detail. The first two steps are performed using the spectral
deconvolution method in the same way as described above in Section~\ref{specdeconv}. The search for
companion objects (third step) is composed of three different steps:
\begin{itemize}
\item For each pixel of the image we compare the flux included in a circle centered on the analyzed pixel and the flux into
an external annulus. The radii values of the circle and of the annulus can be chosen by the user, but for our analysis,
we always adopted the values of 1.5, 2, and 4 (pixels). The user can choose the type of statistic to be performed 
on these regions: a mean or a median. From our tests we find that the second one was more effective in finding companion 
objects, so we always adopted it for all subsequent analysis. The procedure finds an object
if the value found for the inner circle is greater than for the outer annulus plus the standard deviation (on the 
outer annulus) multiplied by a factor that can be chosen by the user and that has to be considered carefully case by case.
\item If finding more than one object into a radius of 3 pixels, the procedure then retains only the most luminous.
\item Finally a two-dimensional Gaussian fit is performed on a small region around the newly discovered object to find its precise
position (in 1/1000 of pixel - no evaluation of the error on the position is done in this procedure). We try to minimze 
the difference between the extracted PSF and the fitting function by performing an iterative procedure to search for the 
minimum of the difference by changing the parameter of the Gaussian fitting 
function.
\end{itemize}
We then extracted the spectrum of the newly found object simply summing the flux of the pixels at a distance less than 1 pixel 
on every subtracted monochromatic image and subtracting from this value the median from the external annulus. We made
the same extraction for two positions at a distance of $\pm \lambda /D$ from the object position along the azimuth (and then 
at the same separation of the found object) to evaluate the spectral noise. Subtracting the mean of these two spectra from 
the object spectrum can then improve the final spectral classification that is the last step of our procedure.\\
To this aim, we compared the output spectra of our simulations with a template spectra grid. We considered T-dwarfs from T0 
to T8 and  L-dwarfs from L0 to L8, with the spectral type L7 replaced by L7.5, because we could not find such a
spectrum in the literature. In addition, we also considered F1V, G0V, K5V, M2V and M8V type star spectra. The data for the 
T-dwarfs template spectra were taken from~\citet{Loo07} for T0, from~\citet{Bur04} for spectra from T1 to T5 and for 
T8 and from~\citet{Bur06} for T6 and T7. The data for the L-dwarfs spectra were taken from~\citet{Tes01}. The stellar
spectra have been taken from the IRTF online Spectral Library 
(\texttt{http://irtfweb.ifa.hawaii.edu/\~{}spex/IRTF\_ Spectral\_ Library/index.html}). \\
The spectral classification was obtained by a cross-correlation (using the IDL routine C\_CORRELATE) between the output
spectrum of each simulation and the template spectra. The spectral type with the highest cross-correlation coefficient is
the one assigned to the simulated planet.

\subsection{Results}\label{res}

\subsubsection{Companion detection}
In Tables~\ref{tabpcgj} and~\ref{tabpcgjh}, we display the numbers and the percentages of found
objects divided according to the spectral type of the input spectra of the simulations (second and third columns) for the Y-J and 
the Y-H modes respectively. In the fourth and in the fifth columns of the same tables, we instead report the numbers and the 
percentages of the spurious objects found with our procedure. It is apparent that we are able to find most of the 
simulated objects both for the Y-J and the Y-H modes, but the method works better in the second case. Moreover, 
the number of spurious objects found is much lower for the Y-H-mode case than in the Y-J-mode case. We do
stress that almost all the simulated objects that we are not able to find in the final image are for the cases at a 
separation of 0.3 arcsec where the background noise from the central star is greater. Indeed, it is $100\%$
complete for companions down to a contrast of $3\times 10^{-7}$ and separations of 0.5 arcsec, while it is  
complete at more than $90\%$ for contrasts other than the worst case with a contrast $3\times 10^{-7}$ and separation 
0.3 arcsec.\\
To confirm this we report in Tables~\ref{tabpccontj} and~\ref{tabpccontjh} the number and the percentage
of those found and of the spurious objects as in Tables~\ref{tabpcgj} and~\ref{tabpcgjh} but, in this case divided
according to the luminosity contrast of the simulated objects. It is apparent from these tables that we are able to find almost
all the simulated objects down to a contrast of $10^{-6}$, while we lose more than $25\%$ of the simulated objects with a 
$3\times 10^{-7}$ luminosity contrast using the Y-J-mode. On the other hand, we are able to find more than $90\%$ of the
objects with a contrast of $3\times 10^{-7}$ using the Y-H-mode.

\begin{table}
\caption{Number and percentage of found objects (F.O.) and of spurious objects (S.O.) subdivided by the simulation input 
spectra for the Y-J-mode case. \label{tabpcgj}}
\begin{center}
\begin{tabular}{c c c c c}
\hline\hline
Sp. Type.    &    F.O.   &   $\%$ F.O.   &    S.O.  &    $\%$ S.O.  \\   \hline
T7     &        55 out of 64    &     85.9       &    7 out of 62      &     11.3     \\   
T2     &        61 out of 64    &     95.3       &   24 out of 85     &     28.2     \\   
L8     &        61 out of 64    &     95.3       &   29 out of 90     &     32.2     \\   
L0     &        60 out of 64    &     93.8       &   31 out of 91     &     34.1     \\   
M2    &        59 out of 64    &     92.1       &   24 out of 83     &     28.9     \\   \hline
\end{tabular}
\end{center}
\end{table}

\begin{table}
\caption{Number and percentage of found objects (F.O.) and of spurious objects (S.O.) subdivided by different
contrasts of the simulated objects for the Y-J-mode case. \label{tabpccontj}}
\begin{center}
\begin{tabular}{c c c c c}
\hline\hline
Contrast    &    F.O.   &   $\%$ F.O.   &    S.O.  &    $\%$ S.O.  \\   \hline
$10^{-5}$                  &        80 out of 80    &     100.0       &    44 out of 124      &     35.4     \\   
$3\times 10^{-6}$     &        80 out of 80    &     100.0       &    26 out of 106      &     24.5     \\   
$10^{-6} $                 &        78 out of 80    &     97.5         &    16 out of 94        &     17.0     \\   
$3\times 10^{-7}$     &        58 out of 80    &     72.5         &    29 out of 87        &     33.3     \\   \hline
\end{tabular}
\end{center}
\end{table}

\begin{table}
\caption{Same as Table \ref{tabpcgj}, but for the Y-H-mode. \label{tabpcgjh}}
\begin{center}
\begin{tabular}{c c c c c}
\hline\hline
Sp. Type.    &    F.O.   &   $\%$ F.O.   &    S.O.  &    $\%$ S.O.  \\   \hline
T7     &        63 out of 64    &     98.4       &    1 out of 64      &     1.6     \\  
T2     &        64 out of 64    &     100.0       &   0 out of 64     &     0.0     \\   
L8     &        61 out of 64    &     95.3       &   12 out of 73     &     16.4     \\ 
L0     &        63 out of 64    &     98.4       &   22 out of 85     &     25.9     \\ 
M2    &        61 out of 64    &     95.3       &   14 out of 75     &     18.7     \\  \hline
\end{tabular}
\end{center}
\end{table}

\begin{table}
\caption{Same as Table~\ref{tabpccontj}, but for the Y-H-mode case. \label{tabpccontjh}}
\begin{center}
\begin{tabular}{c c c c c}
\hline\hline
Contrast    &    F.O.   &   $\%$ F.O.   &    S.O.  &    $\%$ S.O.  \\   \hline
$10^{-5}$                  &        80 out of 80    &     100.0       &    25 out of 105      &     23.8     \\   
$3\times 10^{-6}$     &        79 out of 80    &     98.7         &    14 out of 93        &     15.1     \\   
$10^{-6} $                 &        78 out of 80    &     97.5         &    6 out of 84          &     7.1       \\   
$3\times 10^{-7}$     &        75 out of 80    &     93.7         &    4 out of 79          &     5.1      \\   \hline
\end{tabular}
\end{center}
\end{table}

\begin{figure}
\begin{center}
\includegraphics[width=8.0cm]{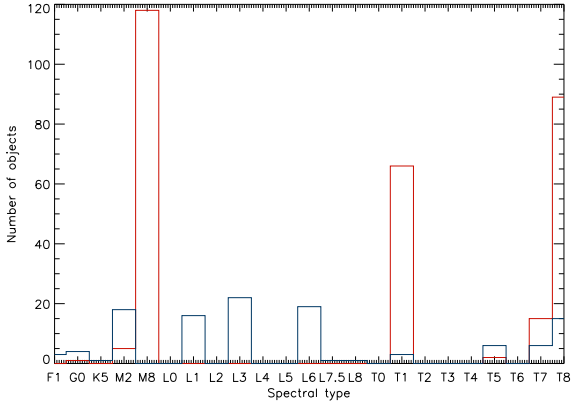}
\caption{Histogram with the number of objects (red) and of spurious objects (blue) found for every spectral type in the
Y-J-mode case. \label{histogram-j}}
\end{center}
\end{figure}

\begin{figure}
\begin{center}
\includegraphics[width=8.0cm]{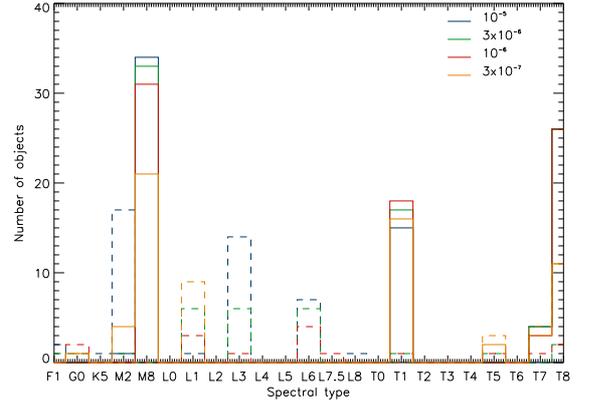}
\caption{Same of Figure~\ref{histogram-j} but with the histograms divided for different luminosity contrast. 
The dashed lines represents the spurious objects. \label{parzhist_j}}
\end{center}
\end{figure}

\begin{figure}
\begin{center}
\includegraphics[width=8.0cm]{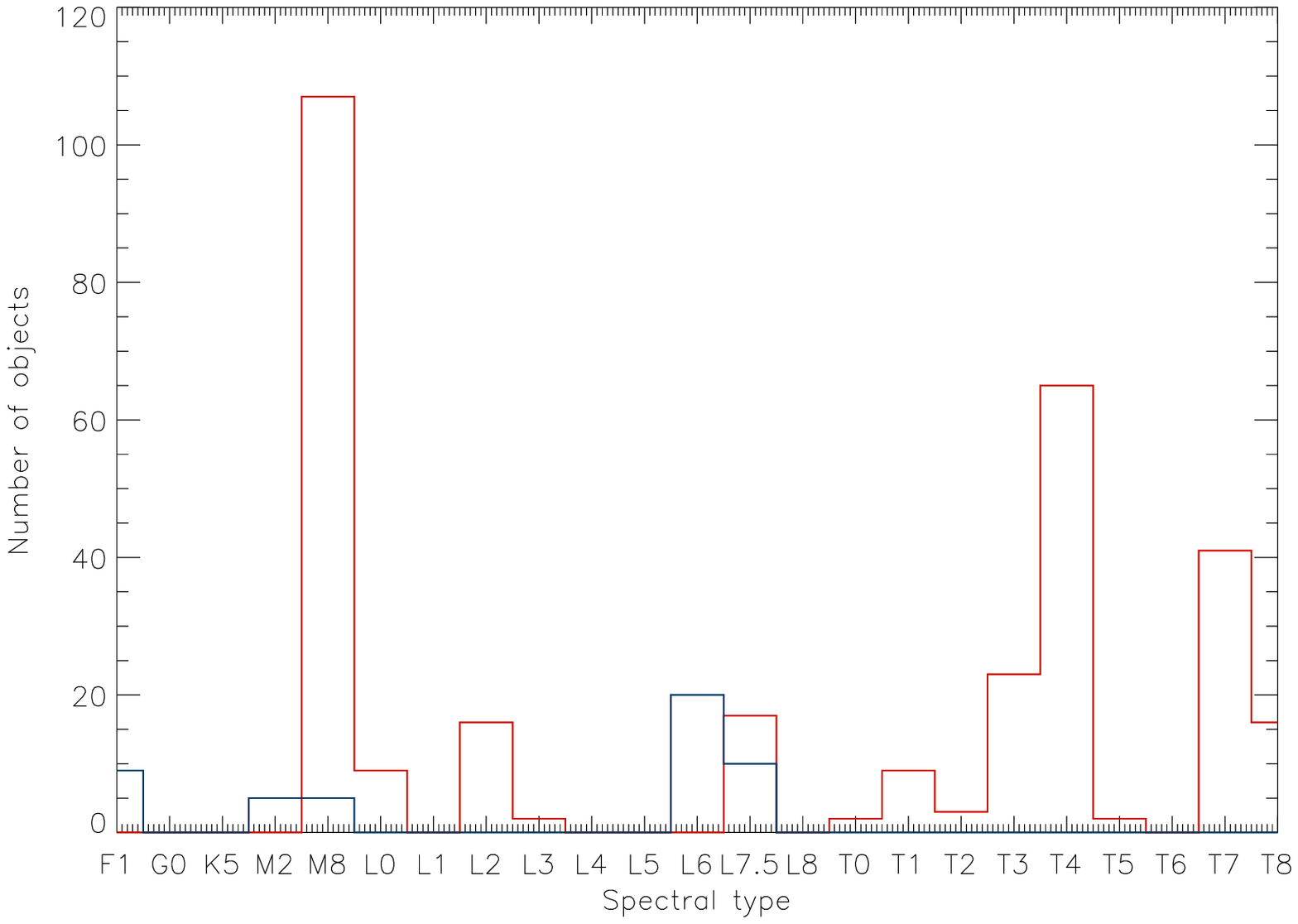}
\caption{Same as Figure~\ref{histogram-j}, but for the Y-H-mode case. \label{histogram-jh}}
\end{center}
\end{figure}

\begin{figure}
\begin{center}
\includegraphics[width=8.0cm]{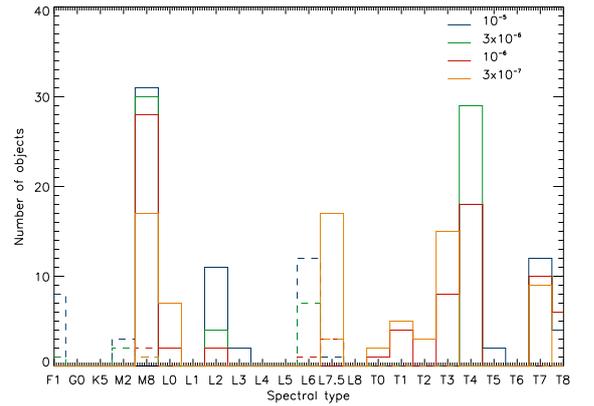}
\caption{Same as Figure~\ref{parzhist_j}, but for the Y-H-mode case. \label{parzhist_jh}}
\end{center}
\end{figure}

\subsubsection{Spectral classification}
In Figure~\ref{histogram-j} we show  the spectral classification of all real objects found, and the spectral classification of the spurious objects. For the real objects, we can see three high peaks corresponding to the M8, T1, and T8 spectral types. The M8 peak is
given by the contribution of objects with both an M2 and an L0 input spectrum. The T1 peak is given by the objects with L8 and
T2 input spectra. In this case, however, the peak is quite low and the objects classification is more dispersed. Finally, the T8
peak is given by T7 input spectra objects. In general, apart from the case of the L0 spectral type, it seems that our procedure
tends to classify the objects with later spectral types rather than the actual ones.  We do not have any particular peak in the 
final distribution for the spurious objects.\\
A similar histogram, but for the Y-H-mode, is displayed in Fig.~\ref{histogram-jh}. Even in this case, we have three peaks 
in the distribution of the real objects. The first one again corresponds to the M8 spectral type and comes from the 
contribution of the simulations with M2 and L0 input spectra objects. This means that, as for the Y-J-mode case, these two 
spectral types seem to originate in a degeneracy. The second peak is around the T4 spectral type and is mainly given by
the T2 input spectra simulations, but from the L8 simulations too. The L8 simulations do not give in general a correct 
identification. Indeed, these objects are recognized alternatively as L2 type or early T type. The last peak is at the T7 spectral
type and it is given exclusively by the T7 simulations objects (the T8 detections are given by the simulations with a 
separation of 0.3 arcsec).
In Figs.~\ref{parzhist_j} and~\ref{parzhist_jh} we display the same histograms, divided in these cases according to the 
different luminosity contrasts of the simulated objects  indicated with different colors in the
figures. From these figures we can see that the overall distribution of the spectral classification is very similar to the 
global one displayed in Figs.~\ref{histogram-j} and~\ref{histogram-jh} down to a contrast of
$10^{-6}$, while they are much more dispersed for the  $3\times 10^{-7}$ contrast where the spectral classification
becomes much less effective.

\subsubsection{Effects of the gravity}\label{gravity}
To further test the capability of our procedure to distinguish different objects, we performed different simulations using as
input the synthetic spectrum of one object with $T_{eff}=800 K$ and $\log (g)=4.0$ and of another one with the same 
temperature and $\log (g)=5.5$. All the simulations were performed for five different objects (with the same characteristics) at a separation from the central star of 0.5 arcsec and a contrast of $3\times 10^{-6}$. Furthermore, we performed 
simulations both for the Y-J and the Y-H-modes.\\
For the simulations with the Y-J-mode, all the objects with $\log (g)=4.0$ were recognized as T8 spectral type (with values of 
the cross-correlation coefficients around 0.75) while the objects with $\log (g)=5.5$ were recognized as T7 (4 cases) or T8 (1 
case). In this second case, the values of the cross-correlation coefficients are on the order of 0.77.\\
On the other hand, for the simulations with the Y-H-mode all the objects with $\log (g)=4.0$ were recognized as T8
spectral type but with higher values of the cross-correlation coefficients (more than 0.93), while all the objects with 
$\log (g)=5.5$ were recognized as T6 spectral type (cross-correlation coefficients on the order of 0.92).\\
In Tables~\ref{tabgravj} and~\ref{tabgravjh} we report the values of the mean coefficients from the cross-correlation between the output and the input spectra for the Y-J-mode and for the Y-H-mode, respectively. From these results it is 
apparent that, in the case of the Y-H-mode we are able to correctly classify the objects for the gravity effects, while for 
the Y-J-mode all the simulated objects are classified as $\log (g)=4.0$, .\\
\begin{table}
\caption{Cross-correlation coefficients considering the effects of the gravity with Y-J-mode. \label{tabgravj}}
\begin{center}
\begin{tabular}{c c c}
\hline\hline
     &     $\log (g)=4.0$   &   $\log (g)=5.5$ \\   \hline
$\log (g)=4.0$   &    0.88      &     0.40    \\  
$\log (g)=5.5$   &    0.80      &     0.51     \\  \hline
\end{tabular}
\end{center}
\end{table}
\begin{table}
\caption{Cross-correlation coefficients considering the effects of the gravity with Y-H-mode. \label{tabgravjh}}
\begin{center}
\begin{tabular}{c c c}
\hline\hline
     &     $\log (g)=4.0$   &   $\log (g)=5.5$ \\   \hline
$\log (g)=4.0$   &    0.89      &     0.77    \\  
$\log (g)=5.5$   &    0.72      &     0.90     \\  \hline
\end{tabular}
\end{center}
\end{table}
In conclusion, from our analysis it seems that the Y-H-mode is the best solution  for correctly distinguishing between objects with different gravities.

\section{Conclusions}\label{conclusion}
We performed detailed simulations of the performances of the SPHERE IFS instrument and considered different
data analysis methods that can be exploited to reduce data coming from the instrument. In 
particular, we exploited the multiple spectral differential imaging (MDI), the spectral deconvolution (SD), and the angular differential imaging (ADI). This latter seems to be especially useful associated with one of the other two 
methods. It turned out that SD is slightly more effective in reducing the speckle noise than the MDI, and it is less sensitive to the
characteristics of the planetary spectrum. From our analysis, however, we can now conclude that, in the best cases, the IFS 
channel of SPHERE should be able to image companion objects around nearby stars down to a contrast of almost $10^{-7}$ 
at a few tenths of an arcsec.\\
We then performed detailed simulations to test the possible impact of Fresnel propagation on the final performances of the 
instrument. This issue created some concerns especially the presence of optics before the lenslet array but, from our 
simulations made under the same IFS optical setup, a negligible difference results between the achievable constrasts with 
or without considering the effects of Fresnel propagation.\\ 
Because the SD method, as said above, allows better results, we used it to perform a new analysis fn the 
capability of the instrument to find and to characterize companion objects of the central star.\\
We then prepared a pipeline with the aim of reducing the datacube resulting from our simulations. To test the
effectiveness of this procedure in finding and characterizing planets, we performed a series of simulations with
different companion objects' input spectra, different separations, and different contrasts between the simulated planets and 
the central star. From these simulations and exploiting the spectral deconvolution method combined with some ADI, we were 
able to image extrasolar planets down to a luminosity contrast with respect to the central star of $3\times 
10^{-7}$. In this way we confirmed the results obtained with the previous run of simulations. \\
We have generally been able to find almost all the simulated objects at the larger separation considered (0.5 and 1.0 arcsec), while
the method is less effective at a separation of 0.3 arcsec. However, even in this case, we were able to find more than the 
$90\%$ of the simulated objects using the Y-J-mode and more than the $95\%$ of the objects using the Y-H-mode.\\
For the spectral reproducibility of our procedure, we can adopt the following conclusions:
\begin{itemize}
\item The greater the separation from the central star, the larger the possibility to reconstruct the planets' spectra with 
precision (considering planets with the same luminosity contrast).
\item Planets with greater luminosity contrast more easily have a precise spectrum reconstruction.
\item This method allows us to reconstruct and to classify the T type spectra very well while spectral reconstruction and 
classification seem to be less precise for earlier spectral types. However, even in these cases, the spectral classification
generally has a precision of a few spectral types (4 or 5 in the worst cases).
\item Stellar spectra (M and earlier spectral types) are clearly distinguished from T-type spectra while some ambiguity is
present for L-type companions. This implies that, in most cases, the characterization of the nature of the detected objects
(companion vs. field star) can be obtained from discovery data alone without waiting for common proper motion confirmation.
Ambigous cases of L-type companions can be disentangled in several cases from the properties of the object and the 
parent star (e.g. L-type companion are expected only above a given contrast threshold).
\item The Y-H-mode allows a better spectral classification than for the Y-J-mode.
\item  For what concerns the effects of the gravity, they are better disentangled using the Y-H-mode than using the Y-J-mode.
\end{itemize}


\begin{acknowledgements}
We wish to thank the referee for the constructive comments on the paper.
SPHERE is an instrument designed and built by a consortium consisting of LAOG, MPIA, LAM, LESIA, Laboratoire Fizeau, INAF, 
Observatoire de Geneve, ETH, NOVA, ONERA, and ASTRON in collaboration with ESO. 
\end{acknowledgements}

\end{document}